# Sub-Hz line width diode lasers by stabilization to vibrationally and thermally compensated ULE Fabry-Perot cavities

J. Alnis[1], A. Matveev[1,2], N. Kolachevsky[1,2], T. Wilken[1], Th. Udem[1], and T.W. Hänsch[1]

[1] *Max-Planck-Institut für Quantenoptik, Hans-Kopfermann-Str. 1, 85748 Garching, Germany*
[2] *P.N. Lebedev Physics Institute, Leninsky pr. 53, 119991 Moscow, Russia*
fax: 0049 89 32905200  email (please do not put online): Janis.Alnis@mpq.mpg.de

We achieved a 0.5 Hz optical beat note line width with ~ 0.1 Hz/s frequency drift at 972 nm between two external cavity diode lasers independently stabilized to two vertically mounted Fabry-Perot (FP) reference cavities. Vertical FP reference cavities are suspended in mid-plane such that the influence of vertical vibrations to the mirror separation is significantly suppressed. This makes the setup virtually immune for vertical vibrations that are more difficult to isolate than the horizontal vibrations. To compensate for thermal drifts the FP spacers are made from Ultra-Low-Expansion (ULE) glass which possesses a zero linear expansion coefficient. A new design using Peltier elements in vacuum allows operation at an optimal temperature where the quadratic temperature expansion of the ULE could be eliminated as well. The measured linear drift of such ULE FP cavity of 63 mHz/s was due to material aging and the residual frequency fluctuations were less than 40 Hz during 16 hours of measurement. Some part of the temperature-caused drift is attributed to the thermal expansion of the mirror coatings. High-frequency thermal fluctuations that cause vibrations of the mirror surfaces limit the stability of a well designed reference cavity. By comparing two similar laser systems we obtain an Allan instability of $2 \times 10^{-15}$ between 0.1 and 10 s averaging time, which is close to the theoretical thermal noise limit.

## 1.    Introduction

Optical atomic clocks and high-resolution laser spectroscopy require spectrally narrow laser light. Even though external cavity diode lasers in general have broad emission spectral line widths on the order of a MHz caused by fast phase fluctuations, their simplicity, ease of use and cost make them attractive if the line width can be reduced significantly. This is accomplished by stabilization to an external high finesse Fabry-Perot (FP) cavity with a narrow resonance line width. To provide sufficient power for frequency conversion to shorter wave lengths master-oscillator power-amplifier (MOPA) systems are used [1-3]. Previously we reported on a MOPA laser system with two frequency doubling stages [4] suitable to excite the narrow 1S-2S two photon transition in atomic hydrogen that posesses a natural linewidth of 1.3 Hz.

In this work we are now reporting on the design of an optical reference cavity for a wavelength of 972 nm that allows the reduction of the laser line width below 1 Hz, while the frequency drift is about 0.1 Hz/s. This wavelength is the 4$^{th}$ sub-harmonic of the 243 nm radiation used to excite 1S-2S two-photon resonance in atomic hydrogen. Besides simplifying a possible optical atomic clock based on 1S-2S transition, this laser system facilitates spectroscopy on more exotic hydrogen-like systems such as tritium, positronium and antihydrogen that are only available in the harsh environment of accelerator laboratories. In future experiments of this type the solid state laser system may replace the dye laser system [5] used so far.

In recent years it has become possible to significantly reduce the laser line widths thanks to improved FP reference resonators. In 2006 a diode laser with 1 Hz line width at 657 nm was reported [6] and in 2007 a diode laser at 698 nm with 0.4 Hz line width was successfully demonstrated [7]. A dye laser possessing sub-Hz line width was already demonstrated in 1999 but with a large and complex vibration isolation setup that consisted of an optical bench suspended on rubber bands [8]. This type of vibration isolation very effectively damps horizontal accelerations due to its swaying motion. In our setup we use a compact active isolation platform that can isolate both horizontal and vertical acceleration by approximately the same amount. The FP cavity mirrors are optically contacted to the ends of a spacer determining the mirror separation $l$. To set the scale for the required length stability one notes that the relative frequency variations $\Delta\nu/\nu$ are identical to the relative mirror distance variations $\Delta l/l$. To reach a stability of $10^{-15}$ using a typical mirror separation $l$ around 10 cm the maximum tolerable length variations are on the order of one tenth of the proton radius. This is the reason why any deforming forces caused by seismic and/or technical vibrations must be prevented from reaching the cavity.

Suspension of the FP cavity in a vacuum chamber using thin wires allows readily excitable pendular motion and for this reason rigid mounting is preferred. We choose to use a vertically oriented FP reference cavity mounted at the mid-plane so that vertical accelerations of the supporting structure leave the separation between mirrors largely unaltered [7, 9, 10]. As shown in Fig. 1 this largely reduces the sensitivity to vibrations. It is also possible to rigidly mount a horizontally oriented at a carefully chosen position to achieve the same effect [11, 12].

Another issue that concerns the stability for averaging times larger than several seconds is the dimensional stability due to temperature variations. Certain glass ceramics can be made with very low thermal expansion and the one made by *Corning* is called Ultra-Low-Expansion glass (ULE). ULE is a titania-doped silicate glass that has a thermal expansion minimum at some temperature $T_c$ around room temperature. When used as a spacer such material reduces the thermal drift considerably and one is left with the quadratic dependence of the relative length:

$$\Delta\nu/\nu = \Delta l/l \sim 10^{-9}(T - T_c)^2, \qquad (1)$$

where $T$ is FP temperature and the coefficient can be determined from the optical beat note frequencies at several temperatures.

To reduce the quadratic dependence the temperature should be stabilized as well as possible, and ideally around $T_c$. For a cavity stabilized at $T_c$ the average drift can be as low as 3.2 mHz/s and the frequency can remain within 1.5 kHz during one day (J.L. Hall, unpublished data). There is yet another drift due to aging of this material. For the cavity used in [5] it is +60 mHz/s calculated from a time span of several years. Unfortunately, quite often $T_c$ of ULE supplied by *Corning* is below room temperature. This poses a problem because cooling of the vacuum chamber that houses the ULE cavity is more difficult than heating as water condensation on the optics may cause problems when temperatures below +15°C are required.

In the following we describe and compare two identical reference FP cavities with quite different housings. The one, called FP1, is stabilized above room temperature by heating the vacuum system. By observing the FP1 frequency at several temperatures above the room temperature we determined $T_c$ at around 7°C. After this finding we built a second reference cavity (FP2) that could be thermally compensated by cooling it to its $T_c$ of 12.5°C with Peltier coolers in vacuum. It turned out that good temperature stabilization is still important for FP2 as the thermal expansion of the mirror coatings starts to make a significant contribution to the frequency stability.

## 2.    Experimental setup
### 2.1    Design of the mid-plane mounted ULE FP cavities

We choose to use vertically mounted FP reference cavities [7] shown in Fig. 1. The spacer is chosen rather short (77.5 mm) to have good mechanical stability and high resonance frequencies. Together with the holding rim it is made from a single piece of premium grade ULE. The cavity mirror substrates are also made from ULE to avoid stress to the spacer caused by thermal expansion. Unfortunately it was not possible to make the mirrors from the same piece as the spacer. The high-quality mirror coatings have 38 layers with a total thickness of 5 μm. The materials used for coatings are $SiO_2$ and $Ta_2O_5$. One mirror is flat and the other is concave with a 50 cm radius of curvature. With these parameters the transverse modes are not degenerate in order to obtain a well separated $TEM_{00}$ modes and a narrow line width. We use two such FP cavities manufactured by *Advanced Thin Films, CO, USA*.

The FP cavity spacer rests on three Teflon posts that have cuts with reduced diameter to soften the support as shown in Fig. 1. This helps to isolate the cavity from horizontal and high-frequency vertical vibrations. To reduce radial forces in the spacer due to thermal expansion of the support structure the posts are resting in holes of a Zerodur ring that has a smaller thermal expansion coefficient than the vacuum chamber itself. For transportation purpose the cavity is fixed loosely to the posts with screws and there is a protecting Teflon ring on the chamber wall.

**2.2 Cavity FP1 stabilized by heating**

The design of the housing is drawn in Fig. 2. The ULE FP cavity is placed in a custom-made cylindrical vacuum chamber made from commonly used technical aluminum alloy. Al allows more uniform heating and vibration damping than stainless steel. For UHV compatibility the Al parts we treated with a strong sodium base to remove the porous oxide followed by immersion into concentrated nitric acid that produces a dense oxide (recipe by C. Vacconeza, DAFNE, available in WWW). All seals are made from indium. With a 3 l/s ion getter pump we achieve the pressure of $1 \times 10^{-7}$ mbar at 30.5°C. A vacuum stability of $1 \times 10^{-8}$ mbar is necessary to reduce the pressure-induced shift to a 1 Hz level. There is a baffle blocking the direct line of sight between the ion pump and the ULE FP because the radiation from its filament can heat the spacer [7].

An analog temperature servo system with one temperature sensor is used for stabilization of the Al vacuum vessel. Strictly speaking such a servo system can fix the temperature at the location of the sensor only, leaving temperature gradients from coupling to the environment, say by heat radiation. These gradients change with the environment limiting the temperature stability that can be achieved by this method. To reduce this effect it is necessarry to operate near the environment temperature and to put the vacuum vessel in another temperature-controlled Al box. It may seem obvious to use several temperature control systems distributed over the Al box to achieve a homogenous temperature. However analog servo systems for such a case are difficult to realize because they tend to work against each other as there are infinitely many stable operating points with different distribution of heating powers among the servos. It turns out that simple controllers, that can turn the heaters only on and off with a 1 mK hysteresis can be used for that purpose without competing. We use 6 such switching controllers, each with its own section of heating foil to control the temperature of the acoustically sealed Al box surrounding the Al vacuum chamber to 30°C as shown in Fig. 2. We calibrated the six AD590 temperature sensors as they can have a 1°C manufacturing offset. We could not observe any noticeable aging of these sensors at a sub-mK level during a few days of callibration measurement.

One of the problems encountered was the crosstalk of switching spikes that eventually synchronized all the controllers and compromised the stability. To avoid this effect we have installed opto-couplers to transmit the switching signals. To decrease temperature oscillations the output currents of the controller channels can be adjusted individually for on as well as for off states. To further suppress the influence of the heat radiation penetrating the setup from outside it is wrapped in multiple layers of highly reflective aluminum-coated Mylar shields widely used in cryogenics. Good thermal insulation from the environment allows to filter out fast ambient temperature fluctuations and allows a more uniform temperature distribution over the Al box.

The temperature stabilized Al box provides a stable temperature environment for the Al vacuum vessel inside that houses the ULE spacer. The vacuum vessel is stabilized at 30.5°C that is only slightly above the temperature of the Al box. The thermal time constant of the vacuum chamber relative to the Al box was measured to 3 hours. The compact size of the vacuum chamber (15 cm height, 12 cm diameter) facilitates good temperature control. It is equipped with two AD590 temperature sensors: one of them used for regulation and the other for monitoring. Each of them provides a current of 1μA/K that is transformed into a signal of 3 V at $T = 300$ K by a 10 kOhm precision resistor (manuf. Vishay) resulting in a voltage sensitivity of 10 mV/K. In principle a 50 kOhm thermistor could give aproximately 5 times larger voltage sensitivity, but thermistors have a non-linear temperature dependence and need 4-wire connection.

Well designed temperature controllers allow to achieve a sub-100 μK temperature stability during several days, see [13] for a review. We use a custom-made analog proportional-integrating (PI) temperature controller with 5 hour integration time constant. A long integration time is achieved using a bank of ten 22 μF foil capacitors possessing low leak currents. The PI controller should be also temperature stabilized to achieve sub-mK stability of the set temperature.

The heating foils are home-made from 0.3 mm diameter transformer wire fixed to a double-sided tape with a pitch of about 5 mm at the outer surface of the vacuum chamber. It covers all sides of this chamber allowing to adjust the power proportionally to the heated area. The temperature excursions measured with the free running sensor were on average 1 mK and less than 5 mK for acquisition times of 10 h and one week respectively.

## 2.3  Cavity FP2 stabilized by Peltier coolers
### 2.3.1 General idea

In order to be able to cool the ULE cavity in vacuum, while maintaining a highly stable temperature, we have tried a novel approach by placing Peltier coolers inside the vacuum chamber (Fig. 3). We borrowed this idea from hydrogen maser designs whose microwave resonator is stabilized by heating two shields in vacuum. We use two Peltier-coolers to control the temperature of two nested heat shields surrounding the FP cavity mimicking an environment close to $T_c$.

The power that the outer cooled shield recives by heat radiation from the surrounding 35°C vacuum chamber is estimated to 1.5 W from its 0.15 m$^2$ area having 95% reflectivity and 0°C temperature. The radiated power can be reduced by introducing an additional high-reflectivity shield.

One might consider using several Peltier elements per shield which would give a more uniform heating/cooling and simplify mounting of the shields. However, this would be more difficult to troubleshoot in a case of failure and in the first design only one Peltier per shield was used.

### 2.3.2  Construction details

Standard CF150 and CF40 parts have been used for the construction of the vacuum chamber. As shown in Fig. 3 the ULE FP cavity is surrounded by two nested cylindrical heat shields with end covers. The shields are made from commonly used technical grade aluminum alloy with surface treated for vacuum compatibility. The walls of the shields have 5 mm wall thickness.

Manufacturers claim that the Peltier elements are generally vacuum compatible. For testing we have connected them with Capton-isolated wires and after washing in methanol installed into an empty vacuum chamber. After a few days of pumping with a 20 l/s ion pump the pressure reached $10^{-8}$ mbar. After this test we proceeded with the mounting of the shields and ULE FP cavity in the vacuum chamber. Nylon screws and spacers matching the height of Peltier elements were used to fix the shields. For good thermal conductivity we glued Peltier elements with Torr Seal epoxy. A disadvantage of nylon is that it it binds moisture but Teflon does not have enough mechanical strength. An even better material for that purpose would have been Vespel. The finished assembly was baked at 90°C for one week and afterwards the 20 l/s ion pump achieves a pressure of $8 \times 10^{-8}$ mbar.

A single $5 \times 5$ cm$^2$ two-stage Peltier element from *Melcor* cools down the assembly in vacuum to $T_c = 12.5$°C using 2 W of electrical power. The lowest temperature achieved was +2°C using 4W of electrical power limited by the maximum current of our temperature controller of 1 A. We use a double stage Peltier element for the outer shell because of its smaller thermal conductivity. A second smaller ($2 \times 2$ cm$^2$) Peltier element is used to control the inner shield. Two AD590 temperature sensors used for regulation are cemented with Torr Seal in the proximity of each Peltier element, and one additional sensor for diagnostics is glued to the surface of the inner shell far away from the in loop sensor. Capton insulated wire and regular solder was used to connect to a 10 pin vacuum feedthrough.

The vacuum chamber surrounding the two heat shields and the cavity does not need to be temperature controlled and is used as a heat sink reaching +34°C. The vacuum chamber is equipped with the AR coated and tilted glass viewports. Another viewports of the same type are used with the outer shield to block thermal radiation to the ULE FP.

After completion of the cooled cavity setup it turned out that the assembly shows pronounced acoustic pickup. For this reason a sound proof box made from heavy plywood was employed. The box also improves the stability of beam pointing to the ULE FP and in future we plan to slightly prestabilize the temperature in that box.

### 2.3.3. Temperature stability

For temperature stabilisation of the Al shields we use home-made analog PI temperature controllers similar to those used for diode laser temperature stabilization. The temperature excursions of the free-running sensor at the inner shield during 5 days were smaller than 15 mK. Even though this is about 3 times worse than for the heated cavity setup the advantage of operating near $T_c$ increases the long-term frequency stability by 2 orders of magnitude. However, even at $T_c$, the thermal expansion of the mirror coatings requires a very precise temperature control. A few sudden temperature jumps by several mK were observed from the free-running sensor in the first week of operation, probably, as the Peltier element characteristics were stabilizing.

### 2.3.4 Discussion

In the following paragraph we explain why we used two Peltier-controlled shields and not just a single Peltier and several passive heat shields. Both the inner and outer shields of FP2 are stabilized to $T_c = 12.5°C$ using the outer and inner Peltier element. The outer shell serves as a heat sink for the inner shell and the vacuum chamber serves as a heat sink for the outer Peltier element. As the temperature sensors are placed near the corresponding Peltier elements, the area near each Peltier is stabilized within 1 mK. For the outer shell the heat radiation is strong compared to the thermal conductivity of the shell creating ~ 0.2°C temperature difference between the place near the outer Peltier and the most distant point from it. This temperature difference varies with the change of room temperature by as much as 10 mK/°C. Without active stabilization of the iner shield it would thermalize to the average temperature of the outer shield and follow its variation. Due to the small temperature difference between the two shields of only 0.2°C, the radiation transfer to the inner shell is reduced to ~ 10 mW, so that the inner Peltier element allows for a much more stable temperature over the whole inner shell. The temperature at the most distant point from the Peltier element changes by less than 1 mK for one degree of room temperature change.

One might think that the inner Peltier element could induce short-time instability of the ULE cavity temperature. Even though we have not observed this effect, one might add additional passive heat shield(s) around the FP cavity in future designs.

**2.4    Laser stabilization**

Both FP cavity assemblies and the necessary optics are mounted on two $40 \times 40$ cm$^2$ active vibration isolation platforms (see Fig. 4.). One is made by *Halcyonics* and the other by *HWL*, both of which provide suitable vibration insulation. In comparison to our previous 486 nm cavity setup mounted on a $1 \times 2$ m$^2$ platform with a total weight of a few hundreds of kilograms [5], the new configuration is significantly more compact and can easily be transported.

The optical setup for a beat note measurement between the MOPA system designed for hydrogen spectroscopy (based on a *Toptica* laser system) and another home-made external cavity diode laser is shown in Fig. 5. The 972 nm radiation is transported to the vibration-isolation platform by a single mode fiber (not polarization maintaining). Fiber frequency fluctuations resulting from acoustic coupling to the environment are actively compensated by an acousto-optical modulator (AOM) [14]. The frequency of the light is shifted by $2 \times 40$ MHz and sent back through the fiber where it makes a beat note with the light reflected from the input end of the fiber cleaved at 0°. The beat note measured by the photodiode PD4 is down-converted to low frequencies using a stable RF synthesizer running at 80 MHz and is used as an error signal for a phase lock loop (PLL) that controls the AOM frequency. The light intensity sent to the FP is stabilised by controlling the AOM driving power using the signal from photodiode PD1.

Each Littrow-type diode laser is actively locked to an independent ULE FP cavity by means of the Pound-Drever-Hall (PDH) locking technique [15]. The PDH lock works like a phase lock between the incoming laser light and the light stored in the FP cavity filtering the phase fluctuations of the laser by acting as a flywheel. For this reason this technique is capable of reducing the line width of a laser diode from the MHz-level to the Hz-level even though the width of the FP transmission peaks may be several kHz. PDH error signal is detected with a Si PIN photodiode (PD2) (*Siemens* BPX66) followed by a NE5211 transimpedance amplifier.

Several effects that can cause drifts of the locking voltage generate by the PDH method signal need to be taken care of in order to reach the ultimate performance and allow for line width reduction from 10 Hz to the sub-Hz domain.

(i) Initially we generated the phase modulation sidebands necessary for the PDH lock via a modulation of the laser diode current at 20 MHz. However the modulated laser radiation was sent through the fiber where it was affected by a temperature sensitive etalon effect which slightly changed the ratio of power of the sidebands. This problem is solved by modulating after the fiber with an electro-optic (phase) modulator (EOM) placed on the vibration isolation platform. The EOM (*Linos* LM 0202) also allows the generation of optimally large sidebands (~10% of power) for obtaining strong PDH error signal. The sidebands created by modulating the laser diode current were smaller than 1% in order not to cause unnecessary power conversion losses in the following frequency doubling (SHG) stages. The EOM frequencies are 16.5 MHz for FP1 and 21.1 MHz for FP2.

(ii) A high-quality crystal polarizer in front of the EOM is adjusted to reduce residual amplitude modulation.

(iii) The EOM temperature is actively stabilized to 0.1 K level to keep the remaining amplitude modulation and the offset in the error signal that it causes constant.

(iiii) It is necessary to use an optical isolator in front of the fiber to remove weak spurious unidentified FP etalon effect. At the same time this isolator is used for effective separation of the fiber-noise beat note.

The PID feedback to the laser diode current has a proportional (P) flat gain spanning from DC to 10 MHz that is adjusted just below feedback loop oscillation. The P branch is combined with two integrators (I) connected in series, each having an adjustable corner frequency from a few kHz to 500 kHz. The double integrator allows to have more feedback gain at lower frequencies. The integrating branch is adjusted for the most reliable locking and the smallest in-loop error signal. A differentiator (D) shunts the integrators and produces a phase advance on the control signal to push the feedback oscillation point to higher frequencies [16]. Due to the limited gain of the high frequency operational amplifiers the integrators are not ideal and have a total DC gain of only 60 dB. Thus the integrators behave like the P branch near DC and allowing to lock the laser frequency only with aproximately 100 Hz repeatability. Reproducible re-locking is achieved when the error signal from fast PID regulator is sent to another slow integrator with large DC gain that controls the piezo-ceramic actuator which acts on the diode laser grating. A commercial fast PID regulator (*Toptica FALC*) for diode laser locking to a high finesse FP cavity became available recently and we obtained a similar laser locking performace as with a home-made regulator.

We typically send ~ 20 µW of laser light into the FP cavity and couple it with 20% efficiency. This minute intensity is chosen to account for the intensity dependence of the FP resonance frequency of ~ 25 Hz/µW of coupled light.

## 3. Measurements

For FP1 the photon lifetime of 33.9(2) µs is measured by the optical ring-down method corresponding to a finesse of 410000. For FP2 the finesse is measured to be 400000. The FSR of 1.931(1) GHz is chosen in such a way, that no higher order transversal modes are present in the vicinity of the $TEM_{00}$ mode. In the experiments described in this section we used the optical beat note between the two diode lasers stabilized to independent cavities as a diagnostic tool.

### 3.1 Zero-expansion temperature

As a first step we have performed a set of measurements to determine the zero expansion temperature $T_c$ of the FP1 and FP2 by measuring the optical beat note frequency while the temperature of one of the cavities is changed (Fig. 6). After each temperature adjustment we allowed for one day of thermalization before performng the next meaurement (the time constant was evaluated to 10 h). From our data we deduce a zero expansion temperature of $T_c$ = 7(2)°C for the heated FP1 cavity and 12.5(1) °C for the FP2 cavity with Peltier cooling. The possible explanation for this difference is the influence of the radial forces from the posts modifying the temperature sensitivity of the whole assembly and shifting the observed zero expansion temperature. Additionally, the mirror substrates are made from a different piece of ULE than the spacers which can cause temperature dependent stress of the spacer and shift the observed $T_c$.

From Fig. 6 we can estimate the advantage of stabilizing the cavity near $T_c$. For the cavity with Peltier coolers the temperature sensitivity is 50 Hz/mK, assuming that $T_c$ is found with 0.1°C accuracy. However, for the cavity heated to 30.5°C with $T_c$ = 7°C the thermal sensitivity is 11 kHz/mK. Somewhat surprisingly, the thermal expansion of the mirror coatings, each having a thickness of only 5 µm and linear CTE of $0.5 \times 10^{-6}$/K, is comparable to the residual thermal expansion of the ULE spacer stabilized within 0.1°C of $T_c$. and having a quadratic CTE of $1 \times 10^{-9}$/K. The estimated drift due to the mirror coatings is 20 Hz/mK. It is necessary to take into account that the mirror coatings deposited on a rather thin substrate have small thermal inertia and change temperature significantly faster than the spacer. Hence even operating exactly at $T_c$, the cavity assembly has a non-zero response to the temperature jumps and requires very stable temperature environment.

To test this effect we deliberately induced cavity drift by near IR radiation from a 30 mW fiber-coupled laser sent in a diverging beam through the top window of the vacuum chamber and heating the ULE cavity. This laser was on for 10 s with a 50% duty cycle and caused an immediate drift of the cavity resonance. The drift rates at different temperatures $T$ were compared. In theory the induced frequency drift amplitude should decrease when $T$ is approaching $T_c$. As expected, the induced drift was lower for FP2 cavity stabilized close to $T_c$ but a small non-vanishing induced-drift still remained even when the cavity temperature was slowly swept across $T_c$. We believe that heating of the mirror coatings of the cavity is responsible for this effect.

### 3.2  Fractional power of the optical carrier

When FP1 was kept at 31°C and FP2 at 12.5°C the optical beat note signal between the two lasers was at ~ 113 MHz. It was analysed by a RF spectrum analyzer. Fig. 7 shows a narrow coherent peak and the high frequency noise sidebands originating from the diode laser. The high frequency noise spanning several MHz is typical for external cavity diode lasers. This noise can be reduced by fast feedback electronics to some extent, but not completely removed. We have checked that the noise from the tapered amplifier can be neglected. The power contained in the carrier was calculated as the area under the beat note peak versus the total area. The results are summarized in Table 1. Typically 99.7(1)% of power in the carrier was achieved with 40 µW of light coupled to the FP cavity. Even if this light power is stabilized to 0.1% it causes 1 Hz line shifts. In terms of stability the optimal power coupled to the FP cavity is approximately 4 µW reducing the power of the carrier to 99.0(1)%, indicating that the high bandwidth PDH photodetector becomes noise-limiting. The lowest coupled power with which the diode laser could be locked to the FP was 0.6 µW.

The noise pedestal of the diode laser may be a problem for the hydrogen experiment where two frequency doubling stages follow the laser and a two-photon processes multiply the sensitivity to phase noise tremendously. Due to this reason the excitation efficiency of hydrogen atoms with the diode laser system [4] is only 40 % compared to the excitation using the dye laser system. In the future we plan to decrease the high frequency noise of the diode laser by increasing the separation between the laser diode and the diffraction grating to approximately 20 cm.

### 3.3.  Sub-Hz resolution

For a more detailed analysis the beat note is down-converted (see Fig. 5) and observed at a high resolution using a Fast-Fourier-Transform (FFT) spectrum analyzer. For this device a 100 Hz span with a 4 s acquisition time was chosen giving a 0.25 Hz frequency resolution. This value is a compromise between the instruments resolution and the drift-caused line broadening. The recorded FFT voltage signals were squared to obtain the power spectral density and then individually fitted by a Lorentzian. The signals were recorded during the time periods when the beat note drift was at minimum. For 13 recorded spectra the full width at half maximum ranges between 0.3 and 0.75 Hz with the average equal to 0.47(14) Hz. For graphical representation (Fig. 8) the individual spectra were centered by frequency, averaged and then fitted by a Lorentzian function.

We found that for troubleshooting it was very convenient to down-convert the beat note signals into acoustic and flash light signals. For this purpose we sent the beat signal to a zero-crossing comparator whose output was connected to a speaker and a miniature lamp allowing to perceive acoustically and visually the amazingly stable difference between the very large laser frequencies.

### 3.4. Drift and the Allan deviation

The beat note between the both stable diode lasers was down-converted to ~ 10 MHz and filtered using a band pass filter with 1 MHz bandwidth. Fig. 9 (a) indicates a sub-Hz laser line width between the two lasers. To distinquish which FP cavity is more stable we measured the absolute drift of each laser using a beat note with a fiber laser based femtosecond frequency comb (*Menlo Systems*) referenced from an active hydrogen maser (Quartzlock CH1-75). These beat notes Fig. 9 (b) and (c) show that the FP2 cavity with Peltier cooling to the zero expansion temperature is much more stable than the FP1 cavity stabilized 23ºC away from the optimal temperature. For the Peltier stabilized cavity the linear aging of the cavity was 63.0(1) mHz/s during 16 hours of continuous measurement. When the linear drift was subtracted and data smoothed within 100 s intervals the residual frequency excursions were less than 40 Hz in 16 hours of measurement. These small frequency excursions might be caused by the temperature changes of the optical setup on the vibration isolation platform.

From the optical beat note drift measurements we construct an Allan deviation which is shown in Fig. 10. Dead-time free counter model FXE (K+K Messtechnik) or SR620 (Stanford Research) is used to measure the beat notes used to calculate Allan deviation. HP53131A counter may not be used for times where the cavity drift is not dominating the measurement because it uses an averaging algorithm for frequency calculation which means a violation of the definition of Allan variance [17]. The data from this counter may not be juxtaposed as this counter has a dead time. The minimum of the measured Allan deviation reaches $2 \times 10^{-15}$ between 0.1 and 10 s while for longer times the drift of the FP1 cavity starts to dominate. High-frequency thermal fluctuations causing vibrations of the mirror surfaces actually set the fundamental limit to the stability of a reference cavity. We calculate the thermal noise floor for a single FP cavity to be $1 \times 10^{-15}$ at 972 nm using the formalism described in [18, 19]. Calculation shows that the flat thermal noise floor decreases with increasing mirror separation. A longer cavity on the other hand deteriorates the ability to control for a homogenous temperature and is more sensitive to mechanical perturbations so that there appears to be an optimum spacer length. Thermal noise decreases also with increased optical mode diameter suggesting to use a more flat cavity mirror, but the beam pointing stability might become a limiting issue.

### 4. Conclusions

We confirm the numerous advantages of the mid-plane mounted vertical FP cavities. Such configuration is virtually immune to vibrations, and one can achieve a sub-Hz level of laser spectral line width in a simple compact setup. We demonstrate that Peltier coolers placed in vacuum allow to operate any ULE FP cavity at its zero expansion temperature $T_c$ at which it is least sensitive to temperature instabilities. The new design of an ULE FP cavity with Peltier elements in vacuum is simpler, gives better performance and needs less assembly time compared to the ULE FP cavity design with heaters. Nevertheless a good controller should keep the temperature constant at the sub-mK level if one aims for sub-Hz stability since the thermal expansion of multilayer mirror coatings starts contributing significantly to the total expansion of the FP cavity in this regime. We have recorded a sub-Hz beat note line width between two 972 nm external cavity diode lasers locked to separate mid-plane mounted ULE FP cavities of similar optical design each having a finesse of $4 \times 10^5$. The Allan deviation of the beat note frequency between two independent stabilized lasers reaches $2 \times 10^{-15}$ at averaging times between 0.1 s and 10 s. The cavities are thermal noise limited as the thermal noise floor for one cavity is calculated to be $1 \times 10^{-15}$. The optical carrier of the stabilized diode laser contains between 99.0% and 99.7% of the total laser power in a 10 MHz span and 20 kHz resolution bandwidth depending on the settings of the servo system.

**ACKNOWLEDGEMENTS**

We thank the Ye/Hall group at JILA for the optical design of the cavities, discussion and training. J.A. acknowledges the support by the Intra-European Marie Curie fellowship. N.K. acknowledges support of the Alexander von Humboldt Foundation, Russian Science Support Foundation and RFBR grants 08-02-00443, 08-07-00127. Work supported in partly by the DFG cluster of excellence Munich Center for Advanced Photonics (MAP).

FIGURE 1. Suspending the Fabry-Perot cavity at the horizontal mid-plane its lower part is stretched while the upper part is compressed by gravitational sag. Choosing the proper mounting point the two effects cancel such that the distance between the mirrors remains unchanged. For the same reason the cavity length is immune to other vertical accelerations caused by seismic and/or technical noise.

FIGURE 2. Assembly of the ULE FP1 cavity which temperature is stabilized by heating. The heaters and the temperature sensors are not shown.

FIGURE 3. Assembly of the ULE FP2 cavity with Peltier coolers in vacuum used for temperature stabilization. The outer shield has AR coated glass windows to block the heat radiation reaching the ULE FP. 4 small holes for pumping the outer shield are not shown.

FIGURE 4. ULE FP optical assembly sitting on a vibration isolation platform. Optical setups of FP1 and FP2 are similar. The light is directed into the FP by a 45° mirror (not shown) reflecting the laser beam upwards. An acousto-optic modulator (AOM) for fiber noise compensation is operated around 40 MHz having part of the light that double-passes this AOM reflected back through the fiber. Light intensity is stabilized by the AOM driving power. PD means photodiode detectors, PDH stands for Pound-Drever-Hall locking error signal. Electro-optic modulator (EOM) frequency is 16.5 MHz for FP1 and 21.1 MHz for FP2.

FIGURE 5. Optical beat note measurement system between two Littrow-type external cavity diode lasers stabilized to FP1 and FP2 resonators. The signals a, b, c and d connect to the corresponding points of Fig. 4 where the acronyms are explained. SHG stands for second harmonic generation. Faraday rotator is used as an optical isolator and also to extract the fiber noise cancellation signal.

FIGURE 6. Determination of the zero-expansion temperature for the FP1 and FP2 cavities (squares and circles respectively) and parabolic fits. Operating an ULE FP cavity near $T_c$ gives a very small temperature sensitivity that is dominated at a short time scale by the thermal expansion properties of the mirror coatings of the order of 20 Hz/mK. CTE is coefficient of thermal expansion.

FIGURE 7. Optical beat note signals recorded using a RF spectrum analyzer at different frequency spans and resolutions allowing to examine the laser noise and the lock quality. 113.8 MHz is subtracted from the frequency axes for convenience.

FIGURE 8. Optical beat note signal at 972 nm representing the short term line width of the two lasers stabilized to the FP1 and FP2 cavities. Inset shows phase stability of the beat note filtered with a 100 Hz low pass filter.

FIGURE 9. (a) Beat note between two 972 nm diode lasers stabilized to FP1 and FP2 cavities. (b) FP1 cavity (stabilized 23°C above its $T_c$) beat with a frequency comb (1 s gate time and 100 s smoothed trace). (c) FP2 cavity cooled to $T_c$ beat with frequency comb (100s smoothed). The frequency excursions are only 40 Hz in 16 hours.

FIGURE 10. The Allan deviation between FP1 and FP2 lasers reaches a minimum of $2 \times 10^{-15}$ for averaging times between 0.1 s and 10 s which is near the calculated thermal noise floor of the cavities. Cavity FP2 kept at $T_c$ is extremely stable as can be seen from the beat note with an optical frequency comb referenced to a hydrogen maser.

TABLE 1. Characterization of the lock quality of the diode laser stabilized to a high-finesse FP cavity at two different light powers injected to the cavity.

| Power coupled to the FP | 40 µW | 4 µW |
|---|---|---|
| Power in carrier | 99.7(1) % | 99.0(1) % |
| Regulation sidebands at | 2.4 MHz | 1.5 MHz |
| Carrier level above noise at 20 kHz resolution | 48 dBm | 42 dBm |

TABLE 1

FIGURE 1

FIGURE 2

FIGURE 3

FIGURE 4

FIGURE 5

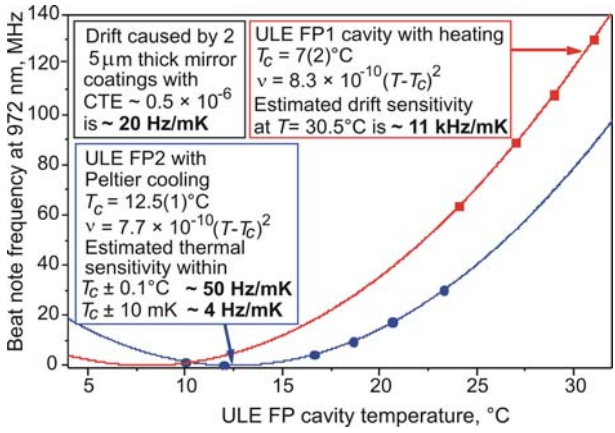

FIGURE 6

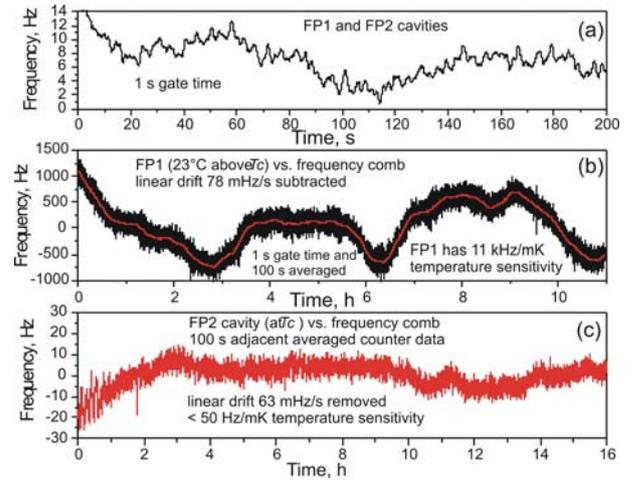

FIGURE 9

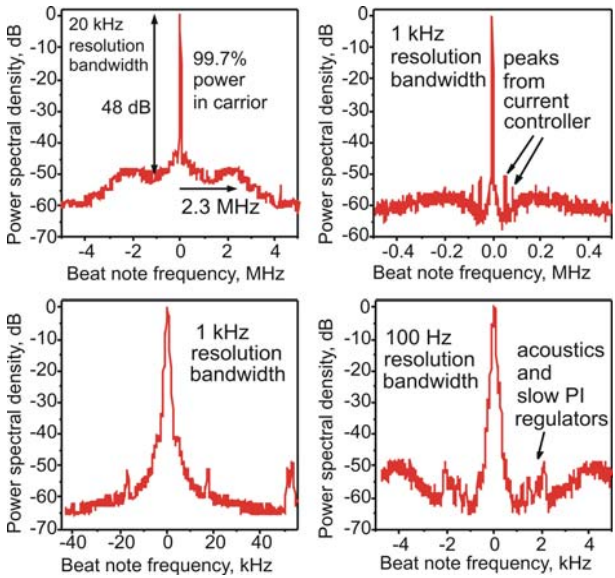

FIGURE 7

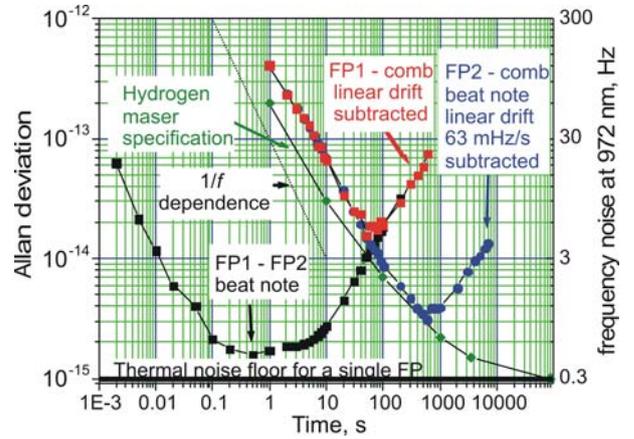

FIGURE 10

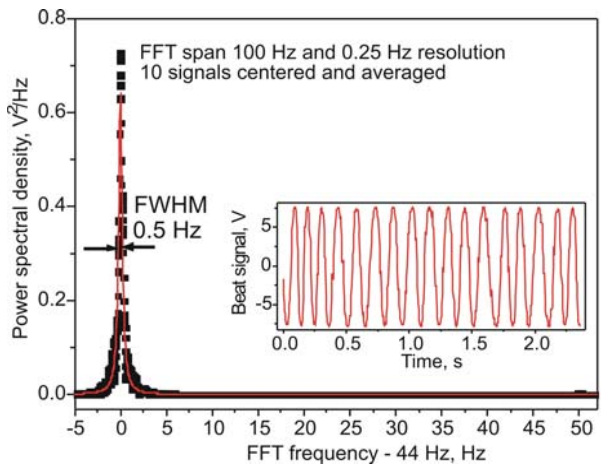

FIGURE 8